\documentclass[aps,prb,twocolumn,groupedaddress,draft,showpacs,intlimits,amsmath,amssymb,floats]{revtex4}
\usepackage{bm}
\usepackage[final]{graphicx}
\usepackage{epsfig}

\begin{document}

\newcommand{\etal}{{\it et al.}\/}
\newcommand{\gtwid}{\mathrel{\raise.3ex\hbox{$>$\kern-.75em\lower1ex\hbox{$\sim$}}}}
\newcommand{\ltwid}{\mathrel{\raise.3ex\hbox{$<$\kern-.75em\lower1ex\hbox{$\sim$}}}}

\title{Collective $d-$wave Excitonic Modes in the Fe-Superconductors}
\author{D.~J.~Scalapino}
\affiliation{Department of Physics, University of California, Santa Barbara, CA 93106-9530 USA}
\author{T.P. Devereaux}
\affiliation{Stanford Institute for Materials and Energy Sciences, SLAC National Accelerator Laboratory, 2575 Sand Hill Road, Menlo Park, CA 94025.}
\affiliation{Geballe Laboratory for Advanced Materials, Departments of Physics and Applied Physics, Stanford University, CA 94305.}

\date{\today}

\begin{abstract}
  
  Calculations of the pairing interaction in multi-band models of the Fe
  superconductors show that it is attractive in both the $A_{1g}$ ($s$-wave)
  and $B_{1g}$ ($d$-wave) channels. This raises the possibility that these
  materials may have collective excitonic modes. Here, assuming an $s$-wave
  groundstate, we investigate the $d$-wave collective excitonic mode and its
  coupling to the Raman scattering.
  
\end{abstract}


\maketitle


Both RPA fluctuation exchange\cite{ref:Kuroki,ref:Graser} and numerical
functional renormalization group calculations\cite{ref:Wang} find that
$s$-wave ($A_{1g}$) and $d$-wave ($B_{1g}$) instabilities can occur
in multi-band models of the Fe-pnictides.
Typically the pairing strength or $T_c$ of the $s$-wave state is found
to be greater than that of the $d$-wave state. Nevertheless, there are
parameter ranges where the two states lay relatively close, raising
the possibility that there could be a $d$-wave ($B_{1g}$) collective
excitonic mode.\cite{ref:Tsuneto,ref:Bardasis} In a fully gapped
superconductor, this mode, consisting of two quasi-particles in a
``Cooper pair" $d$-wave state, lays below $2\Delta$ for zero center
of mass momentum and would appear as a sharp peak. For the Fe superconductors,
calculations suggest that the $s$-wave gap in the groundstate is anisotropic
and may even have nodes. In this case, the collective state could be damped
and appear
as a broad resonance in the two quasi-particle $d$-wave channel. Experimentally
such a mode with $L=2$ could be excited from an $s$-wave superconducting
state by Raman scattering.\cite{ref:rmp,ref:collective,ref:Lee} The observation of such a resonance would be
the first for such a collective state. Early microwave measurements found
indications of a precursor response just below $2\Delta$ in Pb and
it was initially thought that this might be a $p$-wave collective
mode.
However, further studies determined that this was an
artifact.\cite{ref:Martin}  
Here we explore the
possibility of the existence of a $d$-wave collective mode in the Fe-pnictides
and examine how it could be detected by Raman scattering.

\begin{figure}[]
  \includegraphics[width=8cm,clip,angle=0]{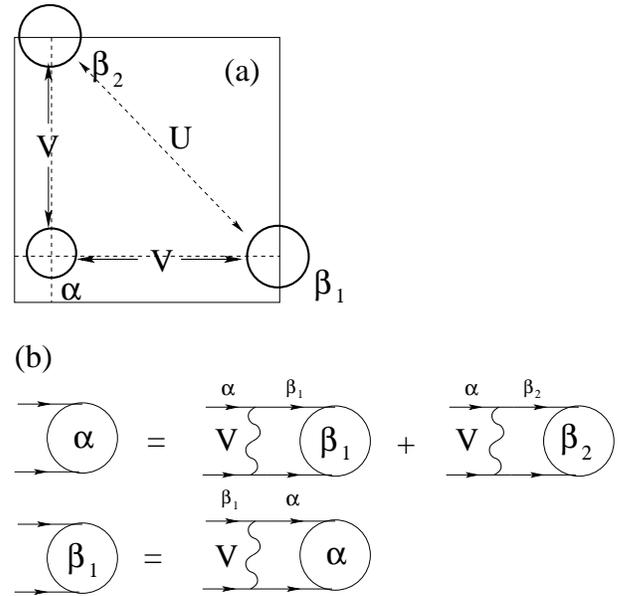}
  \caption{(a) The Fermi surfaces in the 1 Fe/cell Brillouin zone. Here we
    have combined the $\alpha_1$ and $\alpha_2$ hole Fermi surfaces into one
    $\alpha$ sheet. For the simple example that we will consider the pairing
    interaction for the $A_{1g}$ ($s-$wave) ground state is desribed by a strength $V$
    for scattering pairs between the $\alpha$ and $\beta_1$ and $\beta_2$
    Fermi surfaces. The $d-$wave part of the interaction $U$ scatters $d-$wave
    pairs
    between $\beta_1$ and $\beta_2$. (b) Diagrams for the $A_{1g}$ pairing instability.}
  \label{fig:1}
\end{figure}

We will begin by considering the simple case illustrated in Fig.~\ref{fig:1}a.
Here we imagine that the two hole Fermi surfaces around the $\Gamma$ point
of the 1 Fe/cell Brillouin zone have been collapsed into one $\alpha$ Fermi
sheet and $\beta_1$ and $\beta_2$ represent the two electron Fermi sheets.
Suppose the $s$-wave ($A_{1g}$) part of the pairing interaction connecting
the $\alpha$ and $\beta$ Fermi sheets is denoted by $V$ with a cut-off
$\omega_0$ on $\mid\varepsilon_k\mid$ around the Fermi surfaces. There will of
course also be $\beta_1-\beta_2$ contributions to the $s-$wave pair scattering
as well as intra $\alpha, \beta_1$ and $\beta_2$ terms which we neglect in this
simple model. Then the $A_{1g}$
pairing instability represented by the diagrams in Fig.~\ref{fig:1}b is
determined from
\[-2\frac{T_c}{N}\sum_{kn}VG_{\beta_1}(k,i\omega_n)G_{\beta_1}(-k,-i\omega_n)
\phi_{\beta_1}=\phi_\alpha
\]
\begin{equation}
  \mbox{and}
  \label{eq:1}
\end{equation}
\[-\frac{T_c}{N}\sum_{kn}VG_\alpha(k,i\omega_n)G_\alpha(-k,-i\omega_n)
\phi_\alpha=\phi_{\beta_1}
\]
with $G_{\beta,\alpha}(k,i\omega_n)=(i\omega_n-\varepsilon_\beta(k))^{-1}$
the single particle Green's function on the $\beta$ or $\alpha$ Fermi
surfaces, $V$ the $\alpha$--$\beta$ pairing interaction, and $\phi_{\beta,\alpha}$
the gap function amplitude on the $\beta$ and $\alpha$ Fermi surfaces, respectively.
From Eq. (\ref{eq:1}), the transition temperature is determined by
\begin{equation}
  2N_\beta(0)N_\alpha(0)V^2\ln\left(\frac{2\gamma\omega_0}{\pi T_c}\right)=1
  \label{eq:2}
\end{equation}
with the $s$-wave coupling strength $\lambda_s=\sqrt{2N_\beta(0)N\alpha(0)}V$.
The ratio of the gap amplitudes at $T_c$ is
\begin{equation}
  \frac{\phi_\alpha}{\phi_\beta}=-\sqrt{\frac{2N_\beta(0)}{N_\alpha(0)}}
  \label{eq:3}
\end{equation}

In the superconducting state, the dominant $d$-wave scattering between
the quasi-particles occurs between the $\beta_1$ and $\beta_2$ Fermi
surfaces. We will parameterize the $d$-wave part of this interaction by
the separable form
\begin{equation}
  \Gamma^d(k,k')=-g^d_{\beta_1}(\theta)Ug^d_{\beta_2}(\theta')-
  g^d_{\beta_2}(\theta)Ug^d_{\beta_1}(\theta')
  \label{eq:4}
\end{equation}
for $|\varepsilon_k|$ and $|\varepsilon_{k'}|$ less than a cut-off
frequency $\omega_0$. Here, $g^d_{\beta_i}(\theta)$ depends upon the
angle of $k$ on the $\beta_i$-Fermi surface measured from the $k_x$
axis and $g^d_{\beta_2}(\theta)=-g^d_{\beta_1}(\theta+\pi/2)$. 

Similarly to our treatment of the $s-$wave pairing channel, here for
simplicity we will neglect $\beta-\alpha$ contributions to the $d-$wave
channel. The additional contributions to both the $s-$wave and $d-$wave
pairing interactions basically only change the strengths of the effective $s-$
and $d$-wave pairing interactions $\lambda_s$ and $\lambda_d$ which we take as
parameters in the following.

If the system were to remain in the normal state, supercooled below the
$s$-wave pairing instability, it would become unstable to pairing in
the $d$-wave channel when
\begin{eqnarray}
&&  1=\frac{U}{N}\sum_kg^d_{\beta}(k)^2\tanh\beta_c\varepsilon(k)/2\nonumber\\
&&
\simeq
  UN_\beta(0)\int\frac{d\theta}{2\pi}g^d_\beta(\theta)^2
  \ln\left(\frac{2\gamma\omega_0}{\pi T_c}\right).
  \label{eq:5}
\end{eqnarray}
Normalizing the angular average of $g^d_\beta(\theta)^2$ to unity around
the $\beta_1$ Fermi surface, the $d$-wave transition temperature is
\[T_d\sim\omega_0e^{-1/\lambda_d}
\]
with the $d$-wave coupling strength $\lambda_d=N_\beta(0)U$. Fluctuation
exchange calculations\cite{ref:Kuroki,ref:Graser} and numerical
renormalization group studies for models of the Fe superconductors find that the
coupling strength $\lambda_d$ can be comparable to the coupling strength
in the $s$-wave channel $\lambda_s$, raising the possibility that one may
find a $d$-wave ``Cooper Pair" collective mode in the $s$-wave superconducting
state.

The homogeneous Bethe-Salpeter equation for a collective $d$-wave mode is
illustrated in Fig.~2a. Here $k=({\bf k},i\omega_n)$ and $q=({\bf q},i\omega_m)$.
The single- and double-arrow lines denote the single particle Green's
function
\begin{equation}
  G(k)=\frac{i\omega_n+\varepsilon(k)}{(i\omega_n)^2-\varepsilon^2(k)-\Delta^2_{\beta}(k)}
  \label{eq:6}
\end{equation}
and Gor'kov's anomalous Green's function
\begin{equation}
  F(k)=\frac{\Delta_{\beta}(k)}{(i\omega_n)^2-\varepsilon^2(k)-\Delta^2_{\beta}(k)}
  \label{eq:7}
\end{equation}
in the $s$-wave superconducting state, respectively. The Bethe-Salpeter
equation for the $d$-wave collective mode is
\begin{eqnarray}
&&  \phi_q(k)=-\frac{T}{N}\sum_{k',n'}\Gamma^d(k,k')\nonumber\\
&&  \times\left(G(k'+q)G(k')
 +F(k'+q)F(k')\right)\phi_q(k').
  \label{eq:8}
\end{eqnarray}
For the separable interaction given by Eq.~(\ref{eq:4}) one obtains an
equation which determines the energy of the collective $d$-wave mode
\begin{eqnarray}
&& 1=U\frac{T}{N}\sum_{k'n'}\left(g^d_\beta(k')\right)^2\nonumber\\
&&\times  \left[G(k'+q)G(k')+F(k'+q)F(k')\right].
  \label{eq:9}
\end{eqnarray}
After the Matsubara sum is evaluated and $T\rightarrow 0$, we set ${\bf q}=0$
and analytically continue $i\omega_m \rightarrow \omega +i\delta$ to give
\begin{equation}
  1=N_\beta(0)U\int\frac{d\theta}{2\pi}\left(g^d_{\beta_1}(\theta)\right)^2
  \int^{\omega_0}_{-\omega_0}\frac{d\varepsilon}{2}\frac{E}{E^2-\left(\omega/2\right)^2}
  \label{eq:10}
\end{equation}
with $E=\sqrt{\varepsilon^2+\Delta^2_\beta(\theta)}$ and $\omega$ is assumed
to have a small positive imaginary part. Here $\Delta_\beta(\theta)$
is the $s$-wave groundstate gap on the $\beta$ Fermi surface which we will set
equal to $\Delta_0g^s_\beta(\theta)$. The integral over $\varepsilon$ is done
in the usual way
\begin{eqnarray}
&&  \int^{\omega_0}_{-\omega_0}\frac{d\varepsilon}{2}\left(\frac{E}{E^2-(\omega/2)^2}-
  \frac{1}{E}\right)+\int^{\omega_0}_{-\omega_0}\frac{d\varepsilon}{2}\frac{1}{E}\nonumber\\
&&=
  \bar{P}(\omega,\theta)+\ln\left(\frac{2\omega_0}{\Delta_0(\theta)}\right)
  \label{eq:11}
\end{eqnarray}
and extending the range of integration for the first term to plus and minus
infinity gives
\begin{widetext}
\begin{eqnarray}
  \bar{P}(\omega,\theta)&=&
  \frac{\left(\omega/2\Delta_\beta(\theta)\right)}{\sqrt{1-\left(\omega/2\Delta_\beta(\theta)\right)^2}}
  \sin^{-1}\left(\frac{\omega}{2\Delta_\beta(\theta)}\right)\hspace{5mm},\hspace{5mm}
  \frac{\omega}{|2\Delta_\beta(\theta)|}<1 \nonumber\\
  &&\frac{\left(\omega/2\Delta_\beta(\theta)\right)}{\sqrt{\left(\omega/2\Delta_\beta(\theta)\right)^2-1}}
  \left[\ln\left(\left|\frac{\omega}{2\Delta_\beta(\theta)}\right|-
    \sqrt{\left(\frac{\omega}{2\Delta_\beta(\theta)}\right)^2-1}\right)+
    i\frac{\pi}{2}\right],\left|\frac{\omega}{2\Delta_\beta(\theta)}\right|>1. 
  \label{eq:12}
\end{eqnarray}
\end{widetext}

The collective $d$-wave mode at $q=0$ has a frequency and damping given by
\begin{equation}
  \frac{1}{\lambda_d}-\frac{1}{\tilde\lambda_s}=
  \left\langle\left(g^d_\beta(\theta)\right)^2\bar{P}(\omega,\theta)\right\rangle
  \label{eq:13}
\end{equation}
where the bracket implies an angular average. Here the tilde $s$-wave coupling
strength is
\begin{equation}
  \frac{1}{\tilde\lambda_s}=\int\frac{d\theta}{2\pi}
  \ln\left(\frac{2\omega_0}{\Delta_0(\theta)}\right)\left(g^d_\beta(\theta)\right)^2.
  \label{eq:14}
\end{equation}
Depending on the difference in coupling strengths and the anisotropy of the
$s$-wave gap on the $\beta$ Fermi surfaces, one will have a sharp mode or a
resonance.

\begin{figure}[]
  \includegraphics[width=\columnwidth,clip,angle=0]{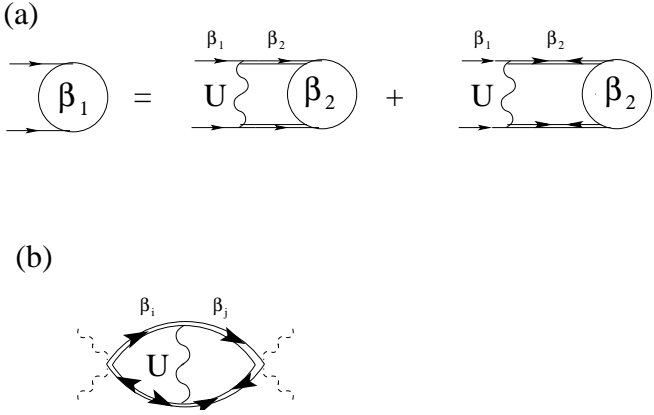}
  \caption{(a) Diagrams for the $d-$wave collective mode. Here the thick
    single arrow line represents the single particle Greens function $G({\bf
      p},i\omega_n)$ and the double arrow line the Gor'kov anomalous Greens
    function $F({\bf p},i\omega_n)$ in the $A_{1g}$ state. (b) The lowest
    order $U$ contribution to the Raman scattering from interband pair
    interactions, scattering pairs between $\beta_1$ and $\beta_2$.}
  \label{fig:2}
\end{figure}

A first order contribution of the interaction vertex $\Gamma^d$ to the Raman
scattering is illustrated in Fig.~\ref{fig:2}b.
There are four arrangements of the $G$ and $F$ propagators and one can go
from $\beta_1$ to $\beta_2$ or $\beta_2$ to $\beta_1$. Finally adding the spin sum,
the first order contribution of $\Gamma^d$ to the Raman susceptibility is
\begin{equation}
  \Delta\chi(i\omega_m)=U(4\gamma_{\beta_1}GFg_{\beta_1}^d)(4g_{\beta_2}GF\gamma_{\beta_2}^d)
  \label{eq:15}
\end{equation}
with
\begin{equation}
  (4\gamma_{\beta}GFg_{\beta}^d)=4\frac{T}{N}\sum_{k,n}\gamma_{\beta}(k)
  G(k+q)F(k)g^d_{\beta}
  \label{eq:16}
\end{equation}
Here $q=(\vec{q},i\omega_m)$ and we are interested in $\vec{q}=0$.
Evaluating the Matsubara sum, we have
\begin{equation}
  (4\gamma_\beta GFg_\beta^d)=N_\beta(0)\Delta_0\langle\gamma_\beta^d(\theta)
  g_\beta^d(\theta)g^s_\beta(\theta)\bar{P}(\omega,\theta)\rangle
  \label{eq:17}
\end{equation}
and
\begin{eqnarray}
\frac{\Delta\chi(\omega)}{N_\beta(0)}=&&-(N_\beta(0)U)
  \left(\frac{\Delta_0}{\omega}\right)^2\nonumber\\
&&\times\langle\gamma_\beta(\theta)
  g_\beta^d(\theta)g^s_\beta(\theta)\bar{P}(\omega,\theta)\rangle^2.
  \label{eq:18}
\end{eqnarray}
Here again the bracket implies an angular average.

\begin{figure}[]
  \includegraphics[width=\columnwidth,clip,angle=0]{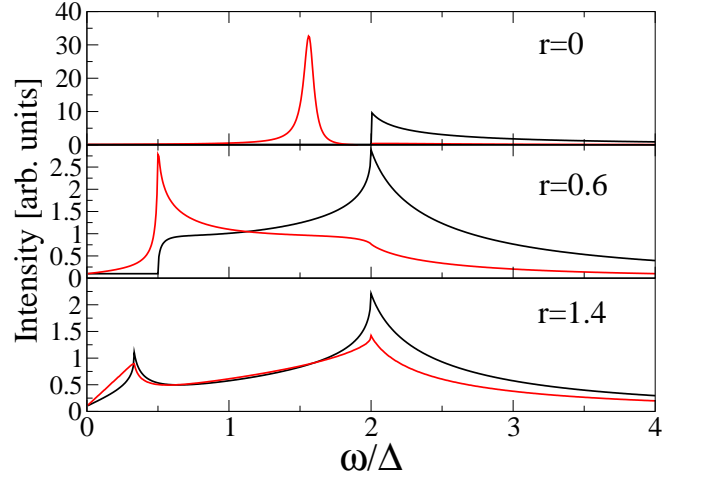}
  \caption{Plots of the Raman response for different gap anisotropies $r$ in the absence of collective mode
    effects (black) and with collective contribution included (red) for
    $\lambda_s=1$ and $\lambda_d=0.5$. Here the parameter $a$ has been set to
    1 and a small damping term has been added. Note the changes in scale of
    the $r=0$ plot.}
  \label{fig:3}
\end{figure}

Symmetry considerations can be used to determine the collective mode
contributions from the interplay of dominant and sub-dominant pair
interactions and polarization geometries. From Eqs. \ref{eq:11}, \ref{eq:12}, \ref{eq:17} and \ref{eq:18}, one can see from symmetry
that the collective mode contribution to the Raman vertex will vanish unless
$g_{\beta}^d\gamma_{\beta}g_{\beta}^s$ transforms as $A_{1g}$ for tetragonal
$D^{4h}$ symmetry. As used here, since $g_{\beta}^sg_{\beta}^d$ transforms as
one of the $d-$wave representation, a collective mode will appear only for
crossed polarization incoming and scattered polarization
geometries. Specifically, if one considers $V$ to be in the $d_{x^{2}-y^{2}}$
channel, the collective mode will appear for $B_{1g}$ orientations
only. 

As previously discussed, multiple scattering between the $\beta_1$ and $\beta_2$ Fermi surfaces leads to
a collective $d$-wave Cooper pair state. Its contribution to the Raman
scattering is obtained by replacing $U$ in Eq.~(\ref{eq:15}) with
\begin{equation}
  \frac{U}{1-N(0)U P(\omega)}
  \label{eq:19}
\end{equation}
Here
\begin{equation}
  P(\omega)=\int\frac{d\theta}{2\pi}\left(g^d_\beta(\theta)\right)^2
  \int^{\omega_0}_{-\omega_0}\frac{d\varepsilon}{2}\frac{E}{E^2-(\omega/2)^2}
  \label{eq:20}
\end{equation}
Proceeding as before, we find that
\begin{equation}
  \frac{\rm{Im}\chi(\omega)}
       {N_\beta(0)}={\rm Im}
       \left\{
       \frac{2\left\langle\gamma_\beta g^d_\beta\left(\frac{2\Delta_\beta(\theta)}{\omega}\right)\bar{P}(\omega,\theta)\right\rangle^2}{\left(\frac{1}{\lambda_d}-\frac{1}{\bar\lambda_s}\right)-\left\langle\left(g^d_\beta\right)^2\bar{P}(\omega,\theta)\right\rangle}\right\}
       \label{eq:21}
\end{equation}

The lowest order contribution to the $\beta$-Fermi surface Raman scattering is given by\cite{ref:paper1}
\begin{equation}
  \frac{\rm{Im}\chi_{\beta\beta}(\omega)}{N_\beta(0)}=\frac{4\pi}{\omega}
  \left\langle\frac{\gamma_{\beta}^2\Delta^2_\beta}{\sqrt{\omega^2-(2\Delta_\beta(\theta))^2}}\right\rangle.
  \label{eq:22}
\end{equation}
If we set $(\gamma_{\beta} g^d_\beta)^2=a\langle\gamma_{\beta}^2\rangle$ in Eq.~(\ref{eq:21})
and (\ref{eq:22}) we have the following
expression for the Raman scattering
\begin{eqnarray}
\frac{\rm{Im}\chi}{N_\beta(0)\langle\gamma_{\beta}^2\rangle}=&&\frac{4\pi}{\omega}
  \left\langle\frac{\Delta^2_\beta}{\sqrt{\omega^2-(2\Delta_\beta)^2}}\right\rangle\nonumber\\
&&+ a{\rm Im}\left\{\frac{2\left\langle\frac{2\Delta_\beta}{\omega}\bar{P}(\omega,\theta)\right\rangle^2}
  {\left(\frac{1}{\lambda_d}-\frac{1}{\bar\lambda_s}\right)-\langle\bar{P}(\omega,\theta)\rangle}\right\}.
  \label{eq:23}
\end{eqnarray}
Here we have also set $\langle g^d_\beta(\theta)^2\rangle=1$. 

\begin{figure}[]
  \includegraphics[width=\columnwidth,clip,angle=0]{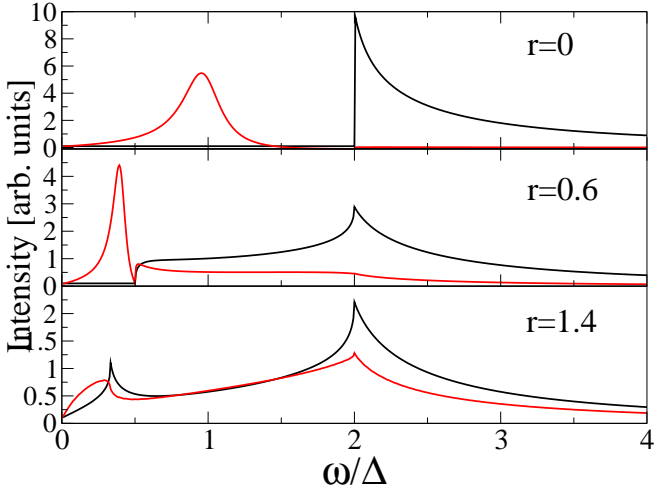}
  \caption{Same as Fig. \ref{fig:3} but for 
    $\lambda_s=1$ and $\lambda_d=0.8$.}
  \label{fig:4}
\end{figure}

Figure~\ref{fig:3} shows plots of $\rm{Im}\chi/N_\beta(0)\langle\gamma_{\beta}^2\rangle$
for $\Delta_{\beta}(\theta)=\Delta_0(1+r\cos2\theta)/(1+r)$ for various values of $r$
and coupling strengths ($\tilde\lambda_s=1, \lambda_d=0.5$).
The results indicate that the collective mode removes spectral
weight from the higher energy portion of the response and adds weight at the
mode position, determined by $1/\lambda_d-1/\tilde\lambda_s$. For the isotropic gap
($r=0$), an essentially undamped mode appears below $2\Delta$, capturing most
of the spectral weight from the bare contribution (the first term in
Eq. \ref{eq:23}). For a gap anisotropy $r=0.6$ that still preserves a
true gap, a well-defined collective mode appears at frequencies slightly below
$2\Delta_{min}$, removing the log singularity of the bare response at
$2\Delta_{max}$ and shifting spectral weight to lower energies. For a gap
with nodes ($r=1.4$), the collective mode is damped by
the finite particle-hole continuum at all Raman energies, and re-allocates
spectral weight to low energies near $\mid2\Delta_{min}\mid$.

For a stronger $d-$wave couplings $\lambda_d=0.8$, appropriate for the case of
near-degenerate pair interactions, the collective mode is pulled further away
from the continuum contribution, as shown in Fig. \ref{fig:4}. For the
isotropic gap case, the mode drops to lower energies and has a smaller
residue. This is the case for $r=0.6$ as well, where a well-defined
collective mode pulls out of the continuum below $2\Delta_{min}$. For the
nodeful case $r=1.4$, the collective mode again drops to lower energies but
remains damped. It however changes the low-frequency behavior of the Raman
response considerably.

For the Fe-pnictides, contributions to the Raman response
comes from each of the $\alpha$ and $\beta_{1,2}$ bands, plus
mixing terms. As the $d$-wave collective
mode contribution arises only from multiple scattering among the $\beta$
bands, the contribution from the $\alpha$ bands will be determined from simple
non-interacting considerations given in Ref. \onlinecite{ref:paper1}. For the case
of degenerate $\beta_{1,2}$ bands having energy gaps 
$\Delta_{\beta_1,\beta_2}(\theta)=\Delta_0(1\pm r\cos2\theta)/(1+r)$,
respectively, the
mixing terms vanish by symmetry for $B_{1g}$ polarizations, and the 
collective mode contribution is twice Eq. \ref{eq:23}, shown in Figs. \ref{fig:3} and \ref{fig:4}.

Since the collective mode contribution is predicted to appear in crossed
initial and scattered photon polarization orientations, this $B_{1g}$
particle-particle mode will not be
coupled to long-range Coulomb forces, which nominally push $A_{1g}$ (parallel
polarizations) collective mode contributions up in energy to the plasma
particle-hole state. This may allow for this mode contribution to appear at low
energies distinctly separate from the continuum in Raman experiments on the
pnictides, unlike the situation in conventional superconductors.\cite{ref:Martin}

Therefore the detailed lineshape of the electronic Raman continuum may be
determined from the interplay of anisotropies of the Raman vertices and
the structure of the pairing interaction using symmetry arguments, as applied successfully in
the cuprates\cite{ref:dvz,ref:rmp,ref:collective}. What may distinguish the Fe pnictides from the
cuprates may be the presence of pairing channels having almost equal strength,
as indicated in recent spin-fluctuation and RG considerations. This paper has
shown that such a circumstance will result in collective mode contribution
which will have an unique polarization signature in the Raman spectrum. This
could open a window into the determination of the pairing structure of the
pnictides and provide important clues to the pairing mechanism.

\section*{Acknowledgement}

We acknowledge important discussions with S. Graser, P. J. Hirschfeld,
T. Maier, R. Hackl, and I. Mazin.
Research was partially supported by DOE DE-AC02-76SF00515
(TPD). DJS thanks the Stanford Institute of Theoretical 
Physics and the Stanford Institute for Materials and
Energy Sciences for their hospitality.

\end{document}